\documentclass[aps,prd,12pt,nofootinbib]{revtex4}

\usepackage{amsfonts}
\usepackage{amsmath,epsfig,bm}
\usepackage{graphicx}
\usepackage{color}

\usepackage[setpagesize=false]{hyperref}

\usepackage{tabularx}  
\newcolumntype{s}{>{\centering\arraybackslash\hsize=.666\hsize}X} 
\newcolumntype{b}{>{\centering\arraybackslash\hsize=1.333\hsize}X}

\usepackage{epsfig}

\newcommand{\be}{\begin{equation}}
\newcommand{\ee}{\end{equation}}
\newcommand{\ba}{\begin{eqnarray}}
\newcommand{\ea}{\end{eqnarray}}
\newcommand{\bd}{\begin{displaymath}}
\newcommand{\ed}{\end{displaymath}}

\begin{document}

\title{Extending a Scaling Equation of State to QCD}

\author{J. I. Kapusta and T. Welle}
\affiliation{School of Physics \& Astronomy, University of Minnesota, Minneapolis, MN 55455,USA}

\begin{abstract}
Whether Quantum Chromodynamics (QCD) exhibits a phase transition at finite temperature and density is an open question.  It is important for hydrodynamic modeling of heavy ion collisions and neutron star mergers.  Lattice QCD simulations have definitively shown that the transition from hadrons to quarks and gluons is a crossover when the baryon chemical potential is zero or small.  We combine the parametric scaling equation of state, usually associated with the 3D Ising model, with a background equation of state based on a smooth crossover from hadrons to quarks and gluons.  Comparison to experimental data from the Beam Energy Scan II at the Relativistic Heavy Ion Collider or in heavy ion experiments at other accelerators may allow the critical exponents and amplitudes in the scaling equation of state to be determined for QCD if a critical point exists.
\end{abstract}

\maketitle

\parindent=20pt

\section{Introduction}

The QCD equation of state has been a subject of much interest and research ever since the discovery of asymptotic freedom.  At very high temperature $T$ and baryon chemical potential $\mu$ it is a weakly interacting gas of quarks and gluons, while at low $T$ and $\mu$ it is a strongly interacting gas of hadrons.  Lattice QCD calculations have shown that the transition from one phase to the other at $T \approx 155$ MeV and $\mu = 0$ is smooth on account of the fact that the up and down quark masses, and consequently the pion mass, are not zero \cite{Aoki2006,Bhattacharya2014}.  However, numerous model calculations predict the existence of a line of first order phase transition terminating in a critical point at $T_c < 155$ MeV \cite{stephanov,CPOD}.   Such a potential critical point is beyond the reach of reliable lattice calculations at this time.  Existing approachess include a Taylor series expansion in powers of $\mu/T$ at $\mu = 0$ \cite{Bazavov2020} and analytic extrapolations from imaginary to real chemical potentials \cite{Borsanyi2020}.  The goal of this paper is to improve upon the constructions of the equation of state reported in Refs. \cite{attract,Taylor1} which is consistent with (i) lattice QCD for all $T$ and small $\mu$, (ii) perturbative QCD for large $T$ and/or large $\mu$, and (iii) a critical point with critical exponents and amplitude ratios from the same universality class as the liquid--gas phase transition and the 3D Ising model.  Parameters in this construction can be adjusted to best fit the experimental data taken during the Beam Energy Scan II at RHIC (Relativistic Heavy Ion Collider) or at other heavy ion accelerators.  

The construction introduced here is an improvement over the ones proposed in Refs. \cite{attract,Taylor1}, which are based on the work of Ref. \cite{Guida} as is ours.  However, our construction has several advantages.  First, our construction is directly in terms of the chemical potential and density.  The approach of Refs. \cite{attract,Taylor1} is in terms of the magnetic field and magnetization in the 3D Ising model.  The mapping from these quantities to the QCD phase diagram introduces significant uncertainty and extra parameters with unknown values which our approach avoids.  Second, in our approach the merging is smooth to all orders, aside from the critical point and its associated line of first order phase transition.  In contrast, Ref. \cite{Taylor1} matched only to a given order of $\mu/T$ in the lattice equation of state Taylor expansion by equating coefficients of the same order.  That can introduce unwanted and/or unphysical phase structures.  Our background equation of state is more sophisticated than that used in Ref. \cite{attract} as is the switching method between critical and background equations of state.

The approach espoused in this paper should be viewed as complementary to that of Ref. \cite{ourletter}.  The approach in that paper is not based on the critical equation of state described in Ref. \cite{Guida} but has the same goal of embedding a line of first order phase transition ending in a critical point in a background equation of state.  

The outline of this paper is as follows.  In Sec. \ref{standard} we summarize the critical equation of state when described in terms of chemical potential and density. 
In Sec. \ref{inclusion} we describe how to combine the critical part with the background equation of state.  In Sec. \ref{params} we discuss the selection of parameters.  In Sec. \ref{results} we show numerical results.  The conclusion is given in Sec. \ref{conclude}.  The appendix contains some comments on closely related work in Refs. \cite{attract,Taylor1}.

\section{Schofield Parametric Scaling Equation of State}
\label{standard}

There is a parameterization of the critical, scaling part of the equation of state which originated more than 50 years ago \cite{Scho1,Scho2,Josephson} and was significantly developed more than 25 years ago \cite{Guida}.  It is
\ba
M &=& m_0 R^{\beta} \theta \nonumber \\
t &=& R (1 -  \theta^2) \nonumber \\
H &=& h_0 R^{\beta \delta} h(\theta)
\ea
where the two independent variables are $R$ and $\theta$.  The $M$ is the magnetization (order parameter), $t = (T - T_c)/T_c$, and $H$ is the magnetic field.  Both $m_0$ and $h_0$ are arbitrary positive normalization constants.  The $h(\theta)$ is an odd function that in the limit $\theta \rightarrow 0$ is normalized so that $h'(0) = 1$.   It must be an analytic function in order to satisfy the requirements of regularity of the equation of state, sometimes referred to as Griffiths’ analyticity.  To obtain the correct ratios of critical amplitudes one usually writes
\be
h(\theta) = \theta ( 1 + h_3 \theta^2 +  h_5 \theta^4)
\ee
and adjusts $h_3$ and $h_5$ accordingly.  If we denote the smallest positive zero of $h$ as $\theta_0$, then the range is $-\theta_0 \le \theta \le \theta_0$.  The second independent variable has the range $R \ge 0$.  The critical exponents are assumed to obey the usual equalities $\alpha + 2 \beta + \gamma = 2$ and 
$\beta (\delta - 1) = \gamma$.  The critical point is at $R = 0$ where $t = M = H = 0$.  In order to describe both $T > T_c$ and $T < T_c$ requires 
$\theta_0 > 1$.  The coexistence curve corresponds to $\theta = \pm \theta_0$, where $H = 0$ and $M_{\pm} = \pm m_0 R^{\beta} \theta_0$.

The conventional discussion and analysis uses Ising model notation and variables.  However, one can use any pair of thermodynamic variables which are conjugate to each other, as pointed out in the reviews \cite{binaryreview,CPRGreview}.  For a liquid--gas transition one may choose the density and chemical potential
\ba
M &\rightarrow& \frac{n - n_c}{n_c} = m_0 R^{\beta} \theta \nonumber \\
H &\rightarrow& \frac{\mu - \mu_c}{\mu_c} = h_0 R^{\beta \delta} h(\theta)
\ea
or the volume per particle and pressure
\ba
M &\rightarrow& \frac{v - v_c}{v_c} = m_0 R^{\beta} \theta \nonumber \\
H &\rightarrow& \frac{P - P_c}{P_c} = h_0 R^{\beta \delta} h(\theta)
\ea
We choose the former.

The pressure must satisfy the condition $(\partial P/\partial \mu)_T = n$.  This can be accomplished by writing
\be
P - P_c = [\mu(R,\theta) - \mu_c] n(R,\theta) - m_0 h_0 \mu_c n_c R^{2 - \alpha} g(\theta)
\ee
where $g$ satisfies the differential equation
\be
(1 - \theta^2) g' + 2(2 - \alpha) \theta g = (1 - \theta^2 + 2 \beta \theta^2) h
\label{geq}
\ee
This is the same function $g$ that appears in Ref. \cite{Guida}.  The constant of integration is fixed by requiring that $g$ be regular at $\theta = 1$.  The solution is
\be
g(\theta) = g_0 + g_1 (1 - \theta^2) + g_2 (1 - \theta^2)^2 + g_3 (1 - \theta^2)^3
\ee
where
\ba
g_0 &=& \frac{\beta ( 1 + h_3 + h_5)}{2-\alpha} \nonumber \\
g_1 &=& \frac{ 1 + h_3 + h_5 - 2 \beta (1 + 2 h_3 + 3 h_5)}{2 (1-\alpha)} \nonumber \\
g_2 &=& \frac{ h_3 + 2 h_5 - 2 \beta ( h_3 + 3 h_5)}{2 \alpha} \nonumber \\
g_3 &=& \frac{(2 \beta - 1) h_5}{2 (1 + \alpha)}  
\ea
The irregular, homogeneous, solution is $(1 - \theta^2)^{2 - \alpha}$; the coefficient is set to zero because it does not contribute to the critical behavior.  The corresponding Helmholtz free energy is
\be
f = \mu_c n + m_0 h_0 \mu_c n_c R^{2 - \alpha} g(\theta) - P_c
\ee

Along the coexistence curve $\theta = \pm \theta_0$ and
\be
\frac{n - n_c}{n_c} = \pm \frac{m_0 \theta_0}{(\theta_0^2 - 1)^{\beta}} (-t)^{\beta}
\ee
Along the critical isotherm $\theta = \pm 1$ and
\be 
\frac{\mu - \mu_c}{\mu_c} = \frac{h_0 h(1)}{m_0^{\delta}} {\rm sgn}(n - n_c) \left| \frac{n - n_c}{n_c} \right|^{\delta}
\ee
These define the critical exponents $\beta$ and $\delta$.

The susceptibility is $(\partial n/\partial \mu)_T = \chi_{\mu\mu}$.  At fixed temperature $(1-\theta^2) dR = 2 R \theta d\theta$.  Thus
\be
\chi_{\mu\mu} = \frac{m_0 n_c}{h_0 \mu_c} \left[ \frac{1-\theta^2 + 2 \beta \theta^2}{(1-\theta^2) h' + 2 \beta \delta \theta h} \right] R^{-\gamma}
\label{sus}
\ee
When the critical point is approached from above ($t \rightarrow 0^+$) at fixed density $n_c$ or chemical potential $\mu_c$, meaning $\theta = 0$, then
\be
\chi_{\mu\mu}^+ = \frac{m_0 n_c}{h_0 \mu_c} t^{-\gamma} \equiv C_+ t^{-\gamma}
\ee
When it is approached along the coexistence curve ($t \rightarrow 0^-$), meaning $\theta = \pm \theta_0$, then 
\be
\chi_{\mu\mu}^- = \frac{m_0 n_c}{h_0 \mu_c} \left[ \frac{1-\theta_0^2 + 2 \beta \theta_0^2}{(1-\theta_0^2) h'(\theta_0) } \right]
(\theta_0^2 - 1)^{\gamma} (-t)^{-\gamma} \equiv C_- (-t)^{-\gamma}
\ee
The critical exponent is $\gamma$ and the ratio of critical amplitudes is
\be
\frac{C_+}{C_-} = \frac{- h'(\theta_0)}{(\theta_0^2 - 1)^{\gamma -1} (1-\theta_0^2 + 2 \beta \theta_0^2)}
\ee

The entropy density is most readily computed from $s = - (\partial f/\partial T)_n$.  At fixed density $\beta \theta dR = - R d\theta$.  Thus
\be
s = \frac{m_0 h_0 \mu_c n_c}{T_c} \tilde{g}(\theta) R^{1-\alpha}
\label{entropyf}
\ee
where
\be
\tilde{g}(\theta) = \frac{\beta \theta g' - (2-\alpha) g}{1 - \theta^2 + 2 \beta \theta^2}
= \frac{g' - h}{2\theta} = \frac{\beta \theta h - (2-\alpha) g}{1-\theta^2}
\ee
Note that the entropy at the critical point is zero in this rendering of the equation of state.  The heat capacity is $c_V = T (\partial s/\partial T)_n$, whence
\be
c_V = \frac{m_0 h_0 \mu_c n_c}{T_c} \left[ \frac{(1-\alpha) \tilde{g} - \beta \theta \tilde{g}'}{1 - \theta^2 + 2 \beta \theta^2} \right] 
\left[ 1 + R (1 - \theta^2) \right] R^{-\alpha}
\ee
When the critical point is approached from above the singular part is
\be
c_V^+ = - \frac{m_0 h_0 \mu_c n_c}{T_c} (2-\alpha) (1-\alpha) g(0) t^{-\alpha} \equiv A_+ t^{-\alpha}
\ee
When it is approached along the coexistence curve the singular part is
\be
c_V^- = - \frac{m_0 h_0 \mu_c n_c}{T_c} \frac{(2-\alpha) (1-\alpha)}{(\theta_0^2 - 1)^{2 -\alpha}} g(\theta_0) t^{-\alpha} \equiv A_- (-t)^{-\alpha}
\ee
The critical exponent is $\alpha$ and the ratio of critical amplitudes is
\be
\frac{A_+}{A_-} = \frac{g(0)}{g(\theta_0)} (\theta_0^2 - 1)^{2-\alpha} 
\ee

Consider the numerical parameters involved.  The critical exponents are universal and taken to be $\beta = 0.3264$ and $\gamma = 1.2371$, resulting in $\alpha \approx 0.1101$ and $\delta \approx 4.7901$ \cite{bootstrap1,bootstrap2}.  The ratios of critical amplitudes are also universal.  The values $h_3 = - 0.762$ and $h_5 = 0.008$, frequently used in the literature, result in $\theta_0 \approx 1.154$, $C_+/C_- = 4.764$, and $A_+/A_- = 0.5302$.  The latter ratios are entirely consistent with published results \cite{Guida,Hasenbusch1,Hasenbusch2}.  For comparison, the mean field approximation has $\alpha = 0$, $\beta = 1/2$, $\gamma = 1$, and 
$\delta = 3$.  In mean field approximation, $C_+/C_- = 2$, and there is a discontinuity in $c_V$ but no divergence. 

Figure 1 shows the coexistence curve in the $T/T_c = 1 + (1-\theta^2) R$ versus $\mu/\mu_c$ plane.  Figure 2 shows isotherms of 
\bd
\frac{P - P_c}{h_0 \mu_c n_c} = h R^{\beta \delta} + m_0 (\theta h - g) R^{2-\alpha}
\ed
versus $n/n_c = 1 + m_0 \theta R^{\beta}$.  This dimensionless plot depends only upon the numerical value of $m_0$.  In order that the density always be positive requires that $m_0 < (\theta_0 ^2 -1)^{\beta}/\theta_0 \approx  0.604$.  For this illustration we chose $m_0 = 0.2$.
\begin{figure}[h]
\includegraphics[trim=30 25 35 35,clip,width=0.8\columnwidth]{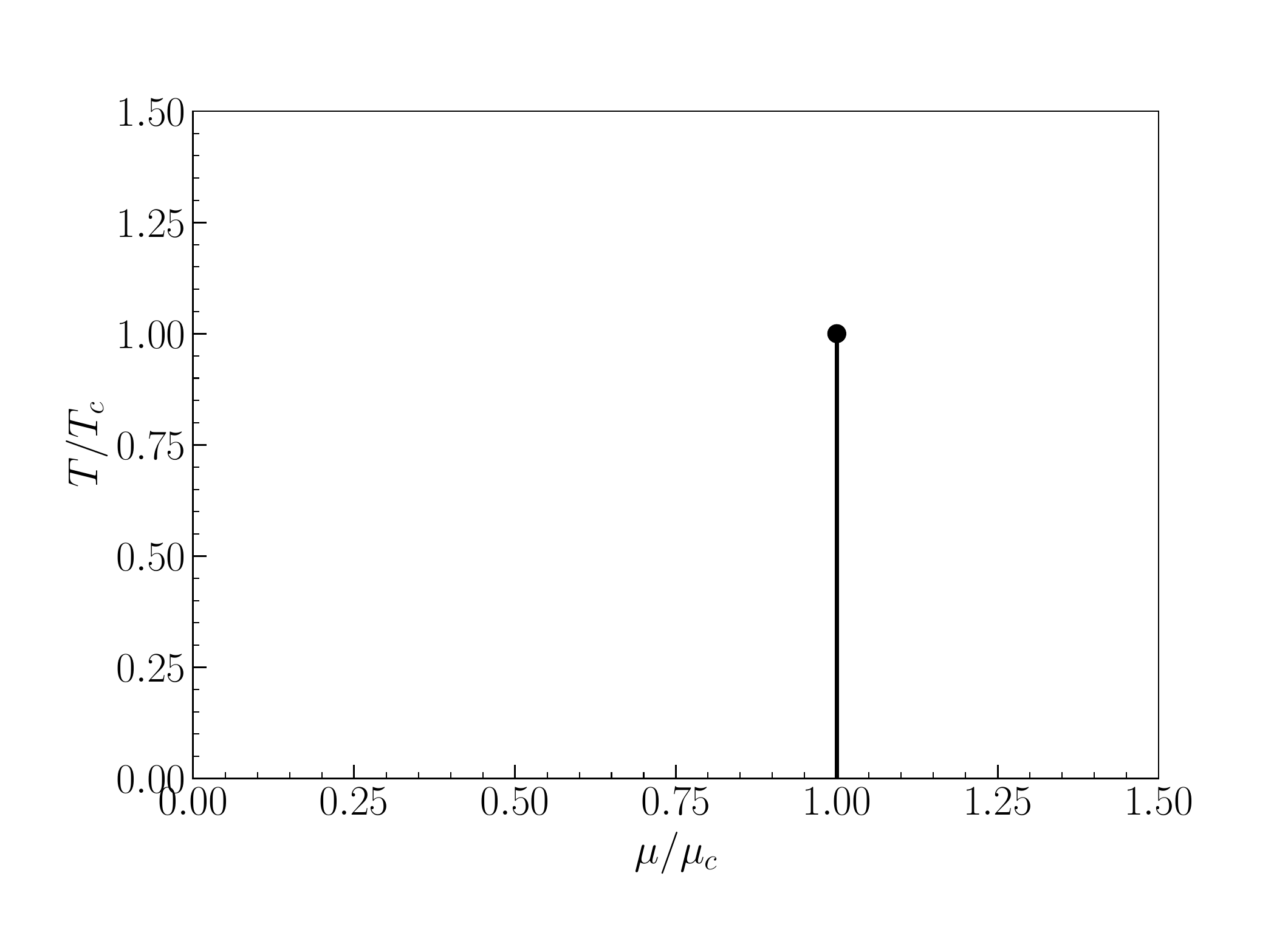}
\caption{The critical curve for the Schofield scaling equation of state. The dot indicates the location of the critical point.}
\label{Fig1}
\end{figure}
\begin{figure}[h]
\includegraphics[trim=30 25 35 40,clip,width=0.8\columnwidth]{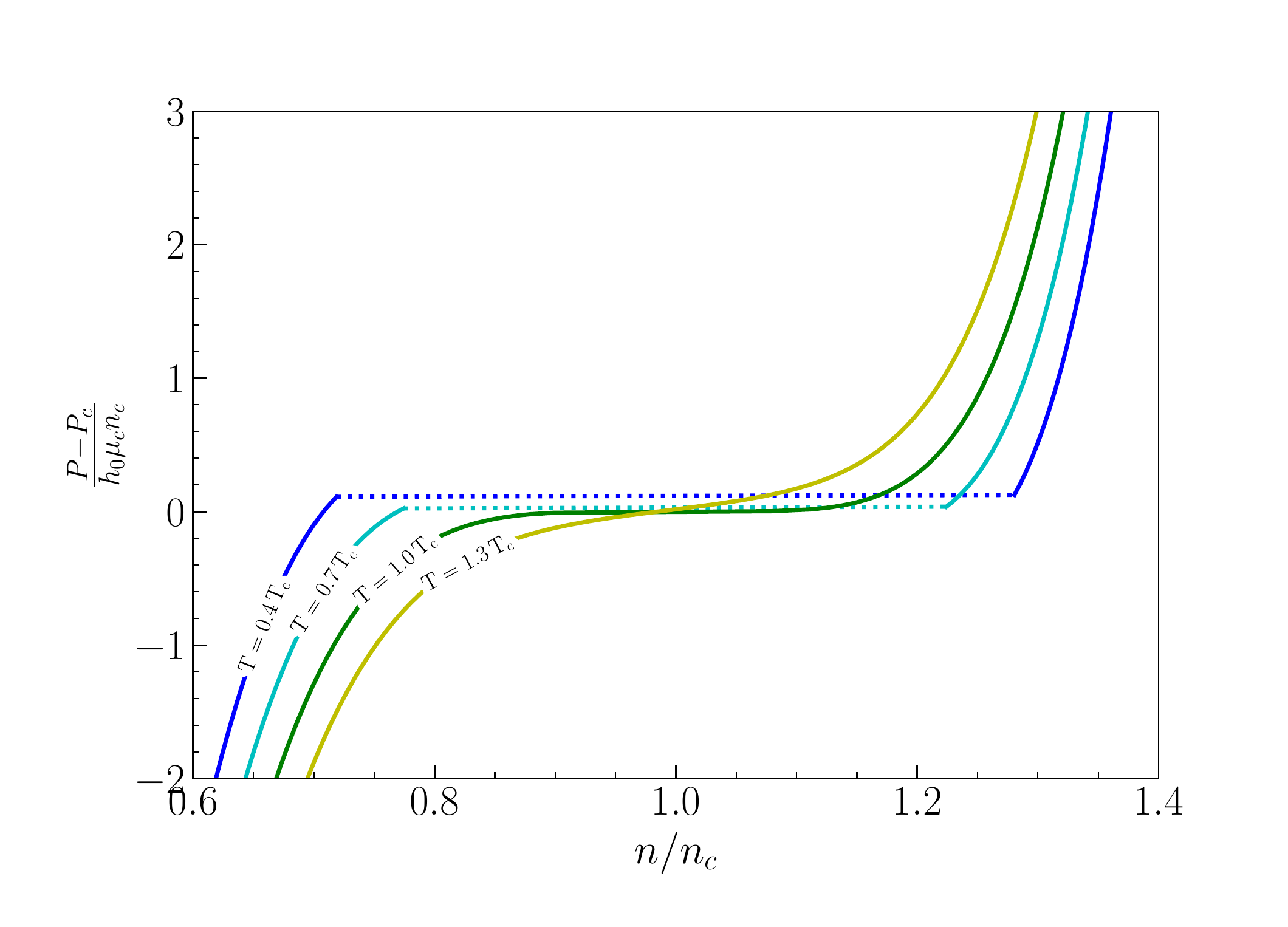}
\caption{Isotherms of pressure versus density for the Schofield scaling equation of state. The only parameter is $m_0 = 0.2$.}
\label{Fig2}
\end{figure}

\section{Including the Background}
\label{inclusion}

The parameterized critical equation of state needs to merge into the background equation of state away from the critical line in the $T$ versus $\mu$ plane.  There is no unique way to accomplish this, and there is no broadly agreed upon method in the literature.  In this section we propose one such method, but it is not the only one.

We express the temperature and chemical potential in terms of the variables $R$ and $\theta$, as in the scaling equation of state, with the explicit inclusion of the coexistence curve described by the function $\mu_x(T)$ by
\ba
\frac{T - T_c}{T_c} = t &=& R (1 -  \theta^2) \nonumber \\
\frac{\mu - \mu_x(T)}{\mu_c} &=& h_0 R^{\beta \delta} h(\theta)
\label{TmuN}
\ea
The first obvious attempt to include the background is to write the pressure as the sum of background plus critical as
\be
P(\mu,T) = P_{BG}(\mu,T) + P_*(R,\theta) 
\ee
where $P_{BG}(\mu,T)$ is a smooth function of $\mu$ and $T$ and
\be
P_*(R,\theta) = P_0 + h_0 \mu_c n_0 R^{\beta \delta} h + m_0 h_0 \mu_c n_0 R^{2 - \alpha} \left[ \theta h(\theta) - g(\theta) \right]
\label{P*}
\ee
is the contribution from the critical part of the equation of state.  The functions $h(\theta)$ and $g(\theta)$ are the same as before.  It differs from the Schofield critical equation of state described in the previous section by the replacements of $P_c$ with $P_0$ and $n_c$ with $n_0$.  The contribution to the pressure from the critical part of the equation of state at the critical point is $P_0$, which could be positive or negative.

Two immediate problems arise: $P_*$ is not an even function of $\mu$, and it generally does not vanish in the vacuum $T = n = 0$.  To address these problems we multiply $P_*$ by the window function
\be
W(\mu,T) = \exp\left[ - \left(\frac{\mu^{2j} - \mu_x^{2j}(T)}{c_* \mu_c^j \mu^j} \right)^2 \right]
\label{Worig}
\ee
where $j$ is a positive integer and $c_*$ is a number which controls the extent of the critical region.  This suppression factor introduces no additional critical behavior.  It goes to zero faster than any finite power of $\mu$ as $\mu \rightarrow 0$ and therefore does not affect any of the susceptibilities at 
$\mu = 0$.  Note that $W(\mu = \mu_x) = 1$ and that $\partial W/\partial \mu (\mu = \mu_x) = 0$.  It is an even function of $\mu$.  So far we have assumed that $\mu \ge 0$ which is the typical situation in heavy ion collisions and neutron stars.  If $\mu < 0$ then one simply changes the signs of $\mu$, $\mu_x(T)$, and $\mu_c$ on the left side of Eq. (\ref{TmuN}).  Hence $P(-\mu,T) = P(\mu,T)$.  In what follows we take $j=1$.

It is worth noting that Ref. \cite{Taylor1} deals with these problems in a different way.  In that approach a Taylor expansion in powers of $\mu/T$ is performed about $\mu = 0$.  Terms up to a finite order are reshuffled between the background lattice equation of state (calculated at $\mu = 0$) and the critical equation of state.  Then a symmetrization is done to ensure that the pressure is an even function of $\mu$.  That procedure limits how large $\mu/T$ can be before unphysical behavior is manifest in the equation of state.  

The pressure is taken to be
\be
P(\mu,T) = P_{BG}(\mu,T) + W(\mu,T) P_*(R,\theta)
\label{pressurerelationN} 
\ee
and so the density is
\be
n = \left( \frac{\partial P}{\partial \mu} \right)_T  = n_{BG}(\mu,T) + W n_* + \frac{\partial W}{\partial \mu} P_*
\label{densityrelationN}
\ee
where
\be
n_* = \left( \frac{\partial P_*}{\partial \mu} \right)_T = n_0 ( 1 + m_0 R^{\beta} \theta)
\label{nstarN}
\ee
Along the coxistence curve $\theta = \pm \theta_0$, $T \le T_c$, and $n_{BG}(T) \equiv n_{BG}(\mu_x(T),T)$.  The critical density is 
$n_c = n_{BG}(\mu_x(T_c),T_c) + n_0$ because $R=0$ at the critical point.  If we want a symmetrical, inverted U shaped curve in the $T$ versus $n$ plane, as approximately seen in the liquid-gas carbon dioxide \cite{binaryreview} and argon \cite{symmetry} phase diagrams, then $\mu_x(T)$ should be determined by the condition $n_{BG}(\mu_x(T),T) = n_c - n_0$ when $T \le T_c$.  In that case the densities along the coexistence curve are
\ba
n_l(T) &=& n_c + m_0 n_0 \theta_0 R^{\beta} \nonumber \\
n_g(T) &=& n_c - m_0 n_0 \theta_0 R^{\beta}
\label{coexistdensitiesN}
\ea
where $n_l$ denotes the high density (liquid) phase and $n_g$ denotes the low density (gas) phase.  Hence the critical behavior is
\be
\frac{n - n_c}{n_0} = \pm \frac{m_0 \theta_0}{(\theta_0^2 - 1)^{\beta}} (-t)^{\beta}
\ee
Along the critical isotherm $\theta = \pm 1$ and
\be 
\frac{\mu - \mu_c}{\mu_c} = \frac{h_0 h(1)}{m_0^{\delta}} {\rm sgn}(n - n_c) \left| \frac{n - n_c}{n_0} \right|^{\delta}
\ee

The susceptibility is $(\partial n/\partial \mu)_T = \chi_{\mu\mu}$.  From Eq. (\ref{densityrelationN}) there are five independent terms.  They are
\be
\chi_{\mu\mu} = W \left( \frac{\partial n_*}{\partial \mu} \right)_T + \cdot\cdot\cdot
\ee
where the first term is the most divergent one near the critical point.  From Eq. (\ref{sus}) this leads to
\be
\chi_{\mu\mu} = W  \frac{m_0 n_0}{h_0 \mu_c} \left[ \frac{1-\theta^2 + 2 \beta \theta^2}{(1-\theta^2) h' + 2 \beta \delta \theta h} \right] 
R^{-\gamma} + \cdot\cdot\cdot
\ee
When the critical point is approached from above ($t \rightarrow 0^+$) at fixed density $n_c$, meaning $\theta = 0$, then the susceptibility diverges as
\be
\chi_{\mu\mu}^+ \rightarrow \frac{m_0 n_0}{h_0 \mu_c} t^{-\gamma} \equiv C_+ t^{-\gamma}
\ee
When it is approached along the coexistence curve ($t \rightarrow 0^-$), meaning $\theta = \pm \theta_0$, then it diverges as
\be
\chi_{\mu\mu}^- \rightarrow \frac{m_0 n_0}{h_0 \mu_c} \left[ \frac{1-\theta_0^2 + 2 \beta \theta_0^2}{(1-\theta_0^2) h'(\theta_0) } \right] 
(\theta_0^2 - 1)^{\gamma} (-t)^{-\gamma} \equiv C_- (-t)^{-\gamma}
\ee
The critical exponent is $\gamma$ and the ratio of critical amplitudes is
\be
\frac{C_+}{C_-} = \frac{- h'(\theta_0)}{(\theta_0^2 - 1)^{\gamma -1} (1-\theta_0^2 + 2 \beta \theta_0^2)}
\ee
This is exactly the same as for the critical equation of state.

The entropy density can be calculated from the pressure as
\be
s = \left( \frac{\partial P}{\partial T} \right)_{\mu} = s_{BG} + W s_* + \frac{\partial W}{\partial T} P_*
\ee
This has no singularities of course.  They arise from higher order derivatives in the critical part of the equation of state.  In this case that means $s_*$.  It can be computed by taking the ratio of
\ba
dP_* &=& h_0 \mu_c n_0 \left[ \beta \delta R^{\beta \delta -1} h dR + R^{\beta \delta} h' d\theta \right] \nonumber \\
&+& m_0 h_0 \mu_c n_0 \left[ (2-\alpha) (\theta h - g) R^{1-\alpha} dR + (\theta h' + h - g') R^{2-\alpha} d\theta \right]
\ea
with
\be
dT = T_c \left[ (1-\theta^2 ) dR - 2 R \theta d\theta \right]
\ee
subject to the condition
\be
d\mu = \mu_x'(T) dT + h_0 \mu_c \left[ \beta \delta h R^{\beta \delta -1} dR + h' R^{\beta \delta} d\theta \right] = 0
\ee
After some algebra, and using Eq. (\ref{geq}), one arrives at
\be
s_* = \frac{m_0 h_0 \mu_c n_c}{T_c} \tilde{g}(\theta) R^{1-\alpha} - \mu_x'(T) n_*
\label{singentropy}
\ee
The singular part of the heat capacity is then computed from $T(\partial s_*/\partial T)_n$.  Now
\be
dn_* = m_0 n_0 (R^{\beta} d\theta + \beta R^{\beta - 1} \theta dR)
\ee
When setting $dn = 0$, with $dR \rightarrow 0$ and $d\theta \rightarrow 0$, the most important term is $dn_*$ as can be seen from the equations for $n$, $dT$, $d\mu$, and $dn_*$.  This means effectively that $n_*$ is constant when taking the derivative.  Hence the results are the same as for the purely critical part of the equation of state, namely, that when the critical point is approached from above, with $\theta = 0$ and $R \rightarrow 0$, the singular part of the heat capacity is
\be
c_V^+ = - \frac{m_0 h_0 \mu_c n_0}{T_c} (2-\alpha) (1-\alpha) g(0) t^{-\alpha} \equiv A_+ t^{-\alpha}
\ee
and when it is approached along the coexistence curve the singular part is
\be
c_V^- = - \frac{m_0 h_0 \mu_c n_0}{T_c} \frac{(2-\alpha) (1-\alpha)}{(\theta_0^2 - 1)^{2 -\alpha}} g(\theta_0) t^{-\alpha} \equiv A_- (-t)^{-\alpha}
\ee
The critical exponent is $\alpha$ and the ratio of critical amplitudes is
\be
\frac{A_+}{A_-} = \frac{g(0)}{g(\theta_0)} (\theta_0^2 - 1)^{2-\alpha} 
\ee

\section{Parameter Selection}
\label{params}

Apart from the choice of the background equation of state there are 7 free parameters: $T_c, \mu_c, P_0, h_0, m_0, n_0, c_*$.  Of course, none of these parameters are universal.  We take $T_c$ and $\mu_c$ as the most interesting and fundamental.

The parameter $P_0$ may be positive or negative, and can be adjusted to produce the desired critical pressure
\be
P_c = P_{BG}(\mu_c,T_c) + P_0
\ee
The parameter $n_0$ must be positive, and can be adjusted to produce the desired critical density
\be
n_c = n_{BG}(\mu_c,T_c) + n_0
\ee
It also determines the critical entropy density as
\be
s_c = s_{BG}(\mu_c,T_c) - \mu_x'(T_c) n_0
\ee
It follows that the critical energy density is
\be
\epsilon_c = -P_c + T_c s_c + \mu_c n_c
\ee

The parameter $m_0$ then determines the strength of the line of first order phase transition.  Along the coexistence curve the baryon density difference is
\be
\Delta n = 2 m_0 n_0 \theta_0 R^{\beta}
\ee
the entropy density difference is
\be
\Delta s = - \mu_x'(T_c) \Delta n
\ee
and the energy density difference is
\be
\Delta \epsilon = [ \mu_x(T) - T \mu_x'(T_c) ] \Delta n
\ee
 
The critical amplitudes for the heat capacity are proportional to $h_0$.  The critical amplitudes for the baryon number susceptibility are inversely proportional to $h_0$.  The reason for the latter is the thermodynamic identity $\chi_B = n/(\partial P(n,T)/\partial n)$ combined with the fact that the piece of the pressure responsible for the critical behavior is proportional to $h_0$ (\ref{P*}).

Finally, the parameter $c_*$ determines the extent of the critical region about the coexistence curve.

\section{Numerical Results}
\label{results}

Any physically reasonable background equation of state may be used.  The approach taken in this paper does not depend on any particular one.  For the sake of illustration, the background equation of state we use involves interpolating between the pressure of a point hadron resonance gas and the pressure obtained from perturbative QCD \cite{Albright}. To interpolate, a switching function $S(T,\mu)$ is used which takes values between 0 and 1 and determines the pressure contributed by each phase. The background pressure is then
\begin{equation}
P_{BG}(T, \mu) = S(T, \mu) P_q(T, \mu) + (1-S(T, \mu))P_h(T, \mu)
\end{equation}
The function $S(T, \mu)$ must be smooth, so as not to introduce unwanted phase transitions, and at $T=\mu=0$ we would like it to approach 0 faster than any finite power of $T$ and/or $\mu$. This is so all derivatives of $S$ vanish at that point, which ensures all thermodynamic observables approach their low energy density values. This function is
\begin{equation}
 S(T, \mu) = \exp \left[-\left(\frac{T^2}{T_s^2} +\frac{\mu^2}{\mu_s^2}\right)^{-2}\right]
\end{equation}
The parameters $T_s=145.33 \; \rm{MeV}$ and $\mu_s=3 \pi T_s$, along with two parameters in the QCD running coupling, are determined by fitting to lattice pressure and trace anomaly results for $100 \le T \le 1000$ MeV and $\mu = 0$. 

From here on we choose $T_c = 120 $ MeV, $\mu_c = 750$ MeV, $P_0 = 0.05 P_c$, $n_0 = 0.1 n_c$, $m_0 = 0.5$, $h_0 = 0.2 $, and $c_* = 0.3$.  For a given critical point $(\mu_c, T_c)$, the critical density is $n_c=n_{BG}(T_c,\mu_c) + n_0$. To calculate the pressure at a particular point $(\mu, T)$, we first calculate $\mu_x(T)$ by solving $n_{BG}(T,\mu_x(T))=n_c - n_0$. This must be done numerically, but isn't difficult. Since $n_{BG}$ is monotonic in both $T$ and $\mu$, it has one solution for each temperature.

\begin{figure}[h]
\includegraphics[trim=30 25 35 40,clip,width=0.8\columnwidth]{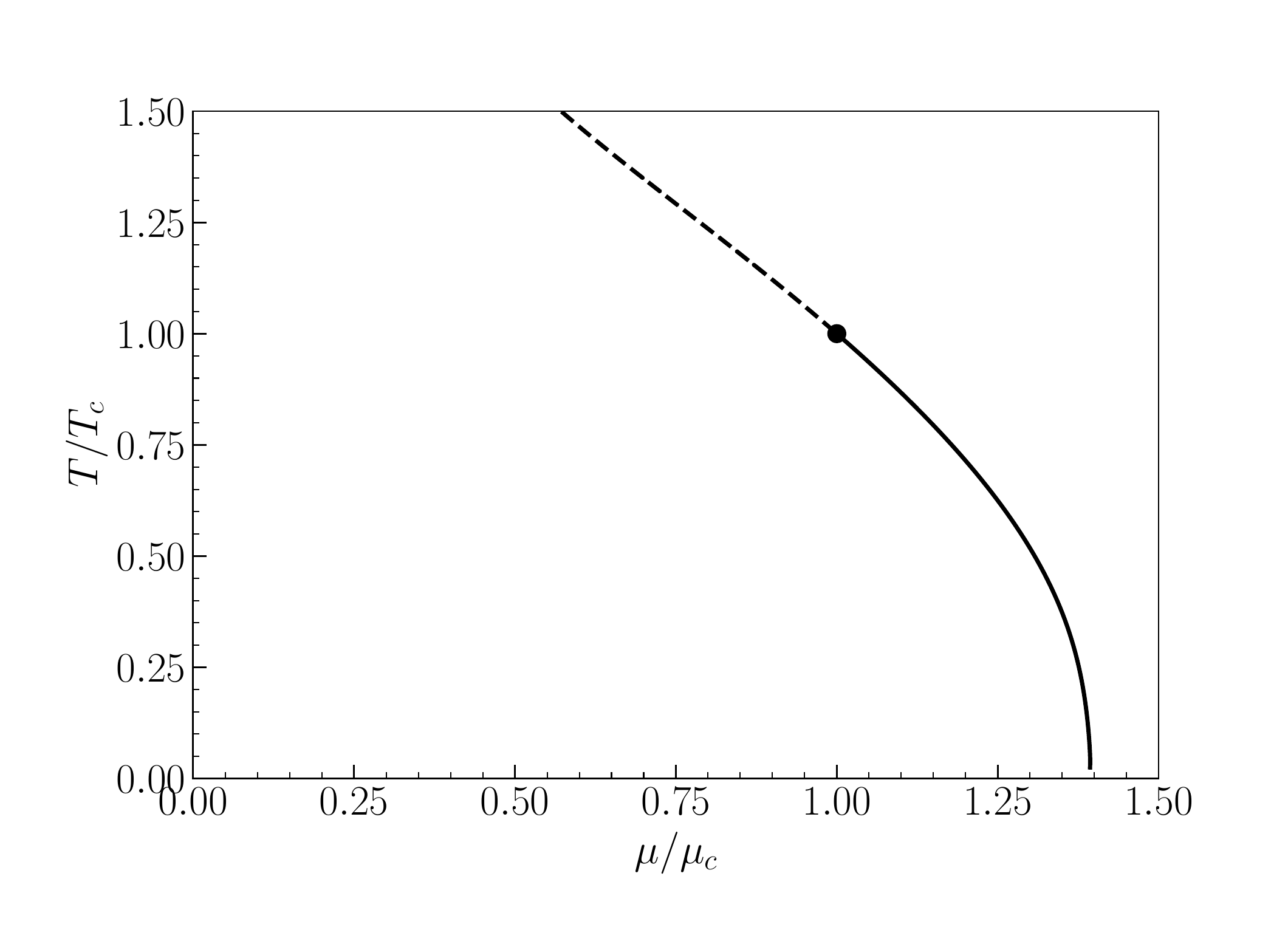}
\caption{The critical curve as described in the text.  The critical temperature is taken to be 120 MeV and the critical chemical potential to be 750 MeV.  The dot indicates the location of the critical point.}
\label{mux}
\end{figure}

The function $\mu_x(T)$ is shown in Fig. \ref{mux}.  As noted in Ref. \cite{ourletter}, this function must be smooth and defined for all temperature, not just for 
$T \le T_c$, in order to avoid unwanted singularities.  The critical density turns out to be $n_c= 1.31 \;\rm{fm}^{-3}$.  To calculate the entropy, we also need the derivative of $\mu_x(T)$, which can be calculated directly form the background via
\begin{equation}
\mu_x^\prime(T) =- \frac{\chi^{BG}_{T \mu}(T,\mu_x(T))}{\chi^{BG}_{\mu \mu}(T,\mu_x(T))}
\end{equation}

\begin{figure}[h]
\includegraphics[trim=0 40 0 55,clip,width=0.80\columnwidth]{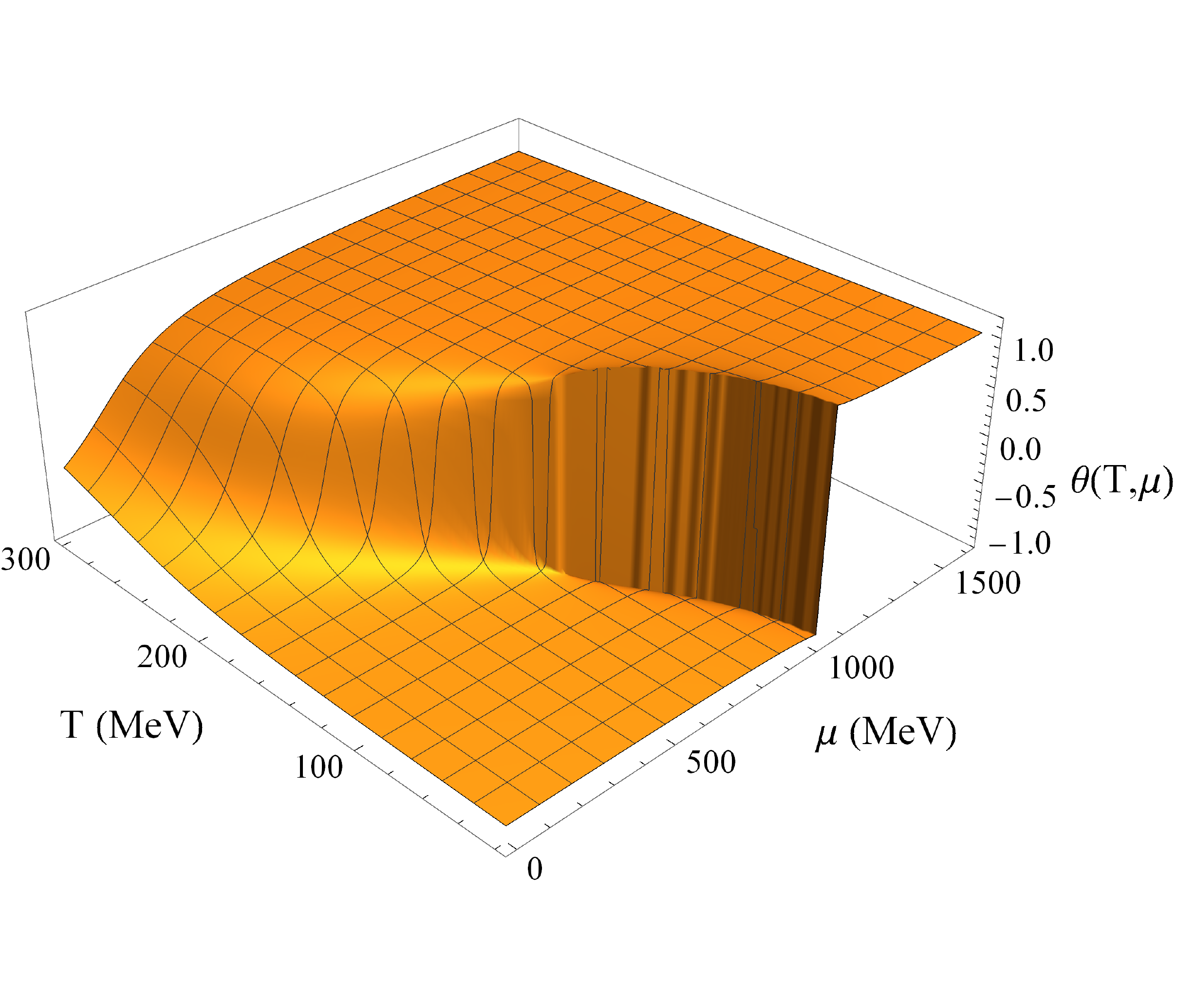}
\vspace{-20pt}
\caption{The function $\theta$ as described in the text.  The critical temperature is taken to be 120 MeV and the critical chemical potential to be 750 MeV.}
\label{theta}
\end{figure}

Now the equation of state can be expressed in terms of $\theta$ and $R$ rather than $T$ and $\mu$. So we must use Eqs. \ref{TmuN} to get these new variables. Once again, this must be done numerically, but the scaling equation is constructed so as to give a unique solution everywhere except along the coexistence curve.  Expressing Eqs. \ref{TmuN} only as a function of $\theta$ we get
\begin{equation}
\frac{h_0 h(\theta)}{\left| 1-\theta^2 \right|^{\beta \delta}} =\left|\frac{T-T_c}{T_c}\right| \left|\frac{\mu-\mu_x(T)}{\mu_c}\right|^{-\beta \delta} {\rm sgn} (\mu-\mu_x(T))
\end{equation}
which can be solved for $\theta$ and then for $R$ while accounting for a number of special cases:

\newpage

\begin{itemize}
\item When $T=T_c$, $\theta= {\rm sgn} (\mu-\mu_c)$
\item When $T>T_c$ and $\mu=\mu_x(T)$, $\theta= 0$
\item When $T<T_c$ and $\mu=\mu_x(T)$, $\theta$ is undefined but goes to $\pm \theta_0$ when approached from above and below the coexistence curve, respectively.
\item $T=T_c$ and $\mu=\mu_c$ is the only point where $\theta=R=0$.
\end{itemize}
The function $\theta(T,\mu)$ is shown in Fig. \ref{theta}.

\begin{figure}[h!]
\includegraphics[trim=36 26 40 30,clip,width=0.75\columnwidth]{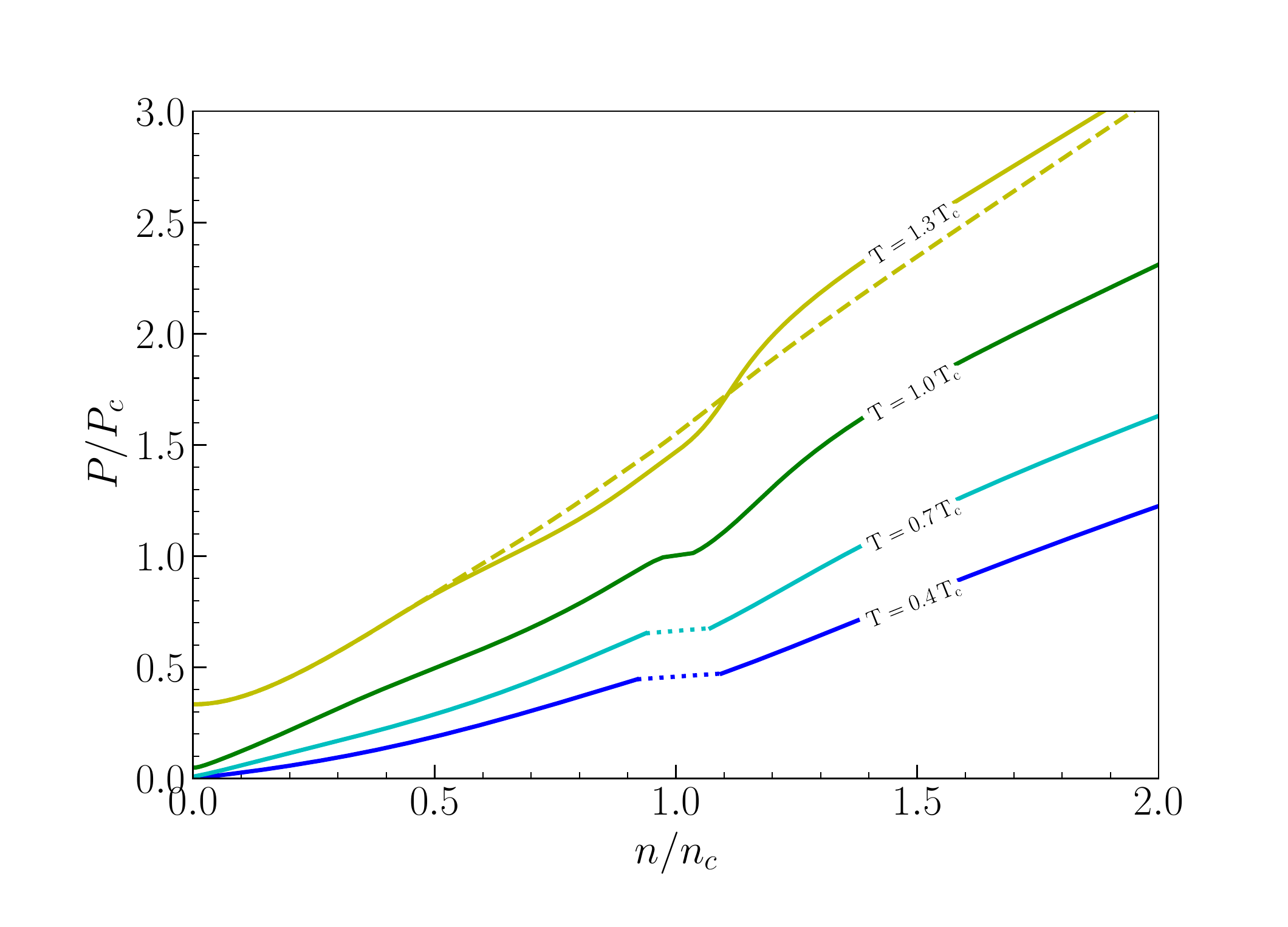}
\vspace{-10pt}
\caption{Isotherms of pressure versus density.  The critical density is $n_c = $1.31 fm$^{-3}$.  The dashed curve is the result of including the factor (\ref{fac}) in the window function. }
\label{P_vs_n}
\end{figure}

Figure \ref{P_vs_n} shows isotherms of pressure versus density.  There is a small residual effect of the critical point above $T_c$.  This is natural.  If desired, this residual effect can be reduced by modifying the window function in such a way that it decreases with temperature when $T > T_c$, not just with distance from the curve $\mu_x(T)$.  The dashed curve is the result of including a factor of
\be
1 - \exp[- (t_0/t)^2]
\label{fac}
\ee
in the window function for $t > 0$ with $t_0 = 0.15$.   

\begin{figure}[h!]
\includegraphics[trim=30 25 40 30,clip,width=0.8\columnwidth]{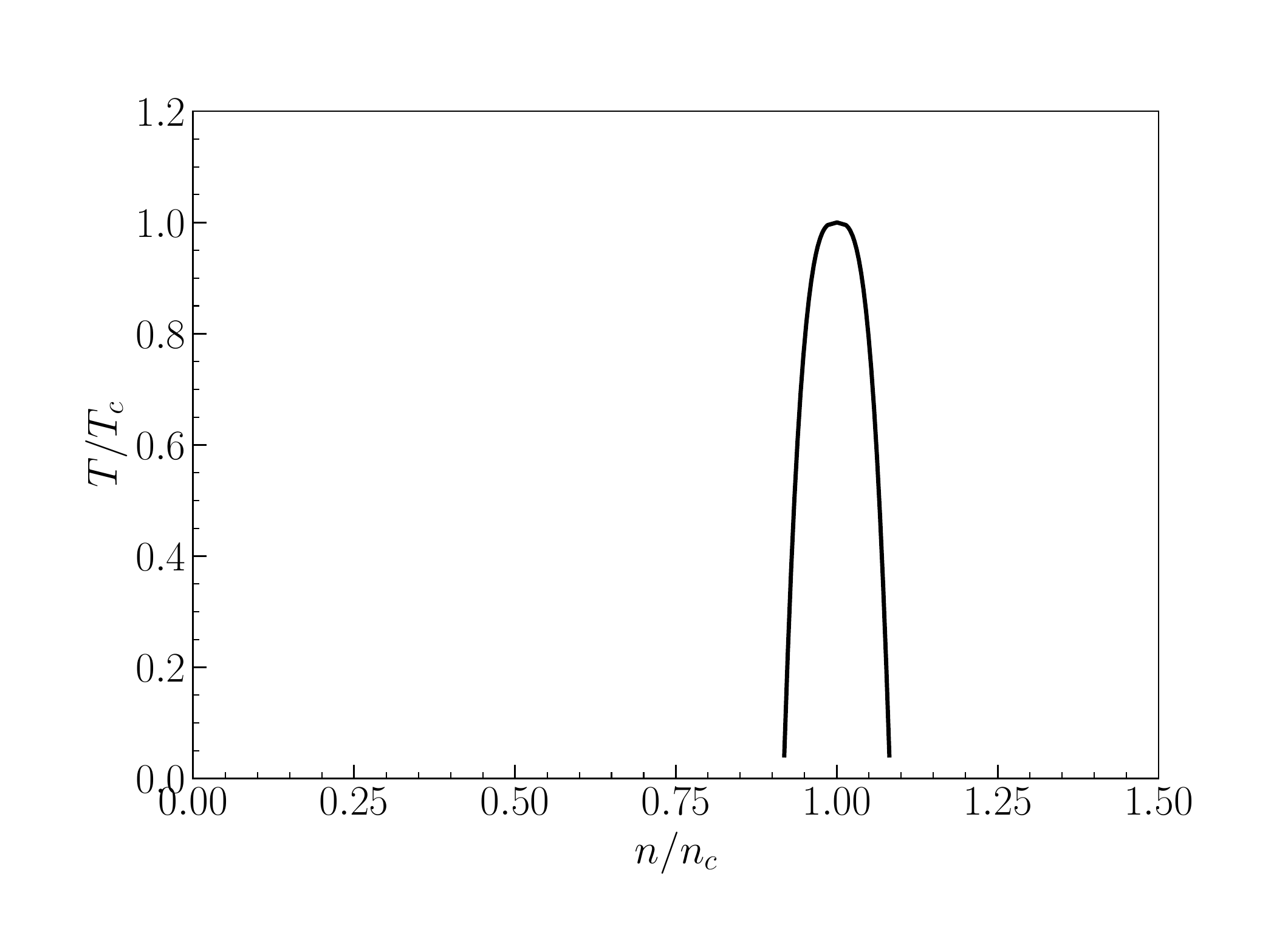}
\vspace{-10pt}
\caption{The coexistence curve as described in the text.  The critical temperature is taken to be 120 MeV and the critical chemical potential to be 750 MeV. The critical density is $n_c = $1.31 fm$^{-3}$.}
\label{T_vs_n}
\end{figure}

Figure \ref{T_vs_n} shows the coexistence curve in the temperature versus density plane.  As discussed in Sec. \ref{inclusion}, it is symmetric about the critical density.

\begin{figure}[h!]
\includegraphics[trim=26 26 40 30,clip,width=0.8\columnwidth]{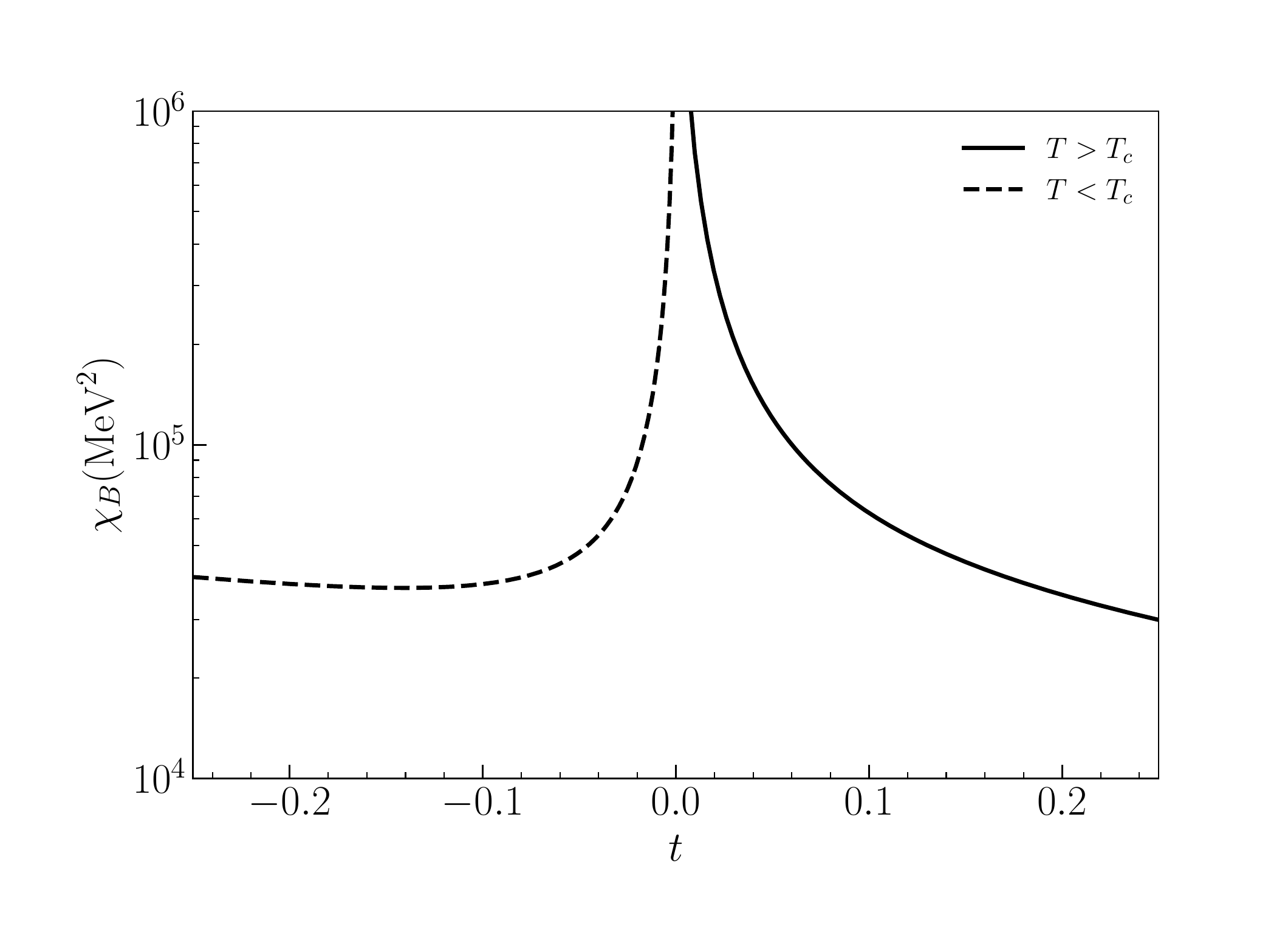}
\vspace{-10pt}
\caption{The baryon number susceptibility as a function of reduced temperature $t$.  The critical temperature is taken to be 120 MeV and the critical chemical potential to be 750 MeV.}
\label{chiB}
\end{figure}

Figure \ref{chiB} shows the baryon number susceptibility as a function of the reduced temperature $t$.  It has the same value in both the high density (liquid) and low density (gas) phases.

\newpage

\vspace{-20pt}

\newpage

\vspace{-20pt}

\newpage

\begin{figure}[h!]
\includegraphics[trim=30 15 50 45,clip,width=0.75\columnwidth]{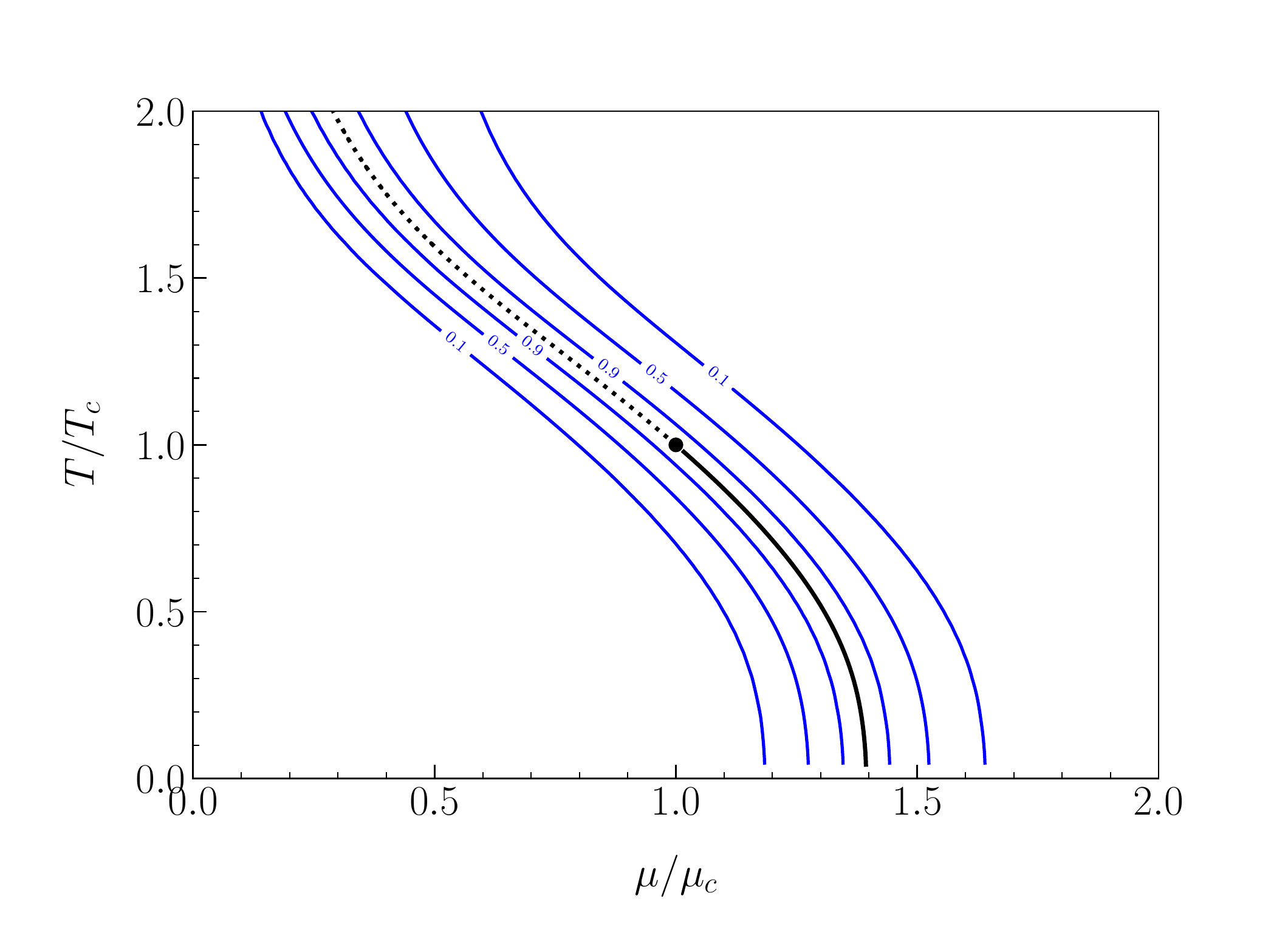}
\vspace{-10pt}
\caption{Window function contours without the factor of (\ref{fac}).}
\label{W1}
\end{figure}

\begin{figure}[h!]
\includegraphics[trim=30 15 50 45,clip,width=0.75\columnwidth]{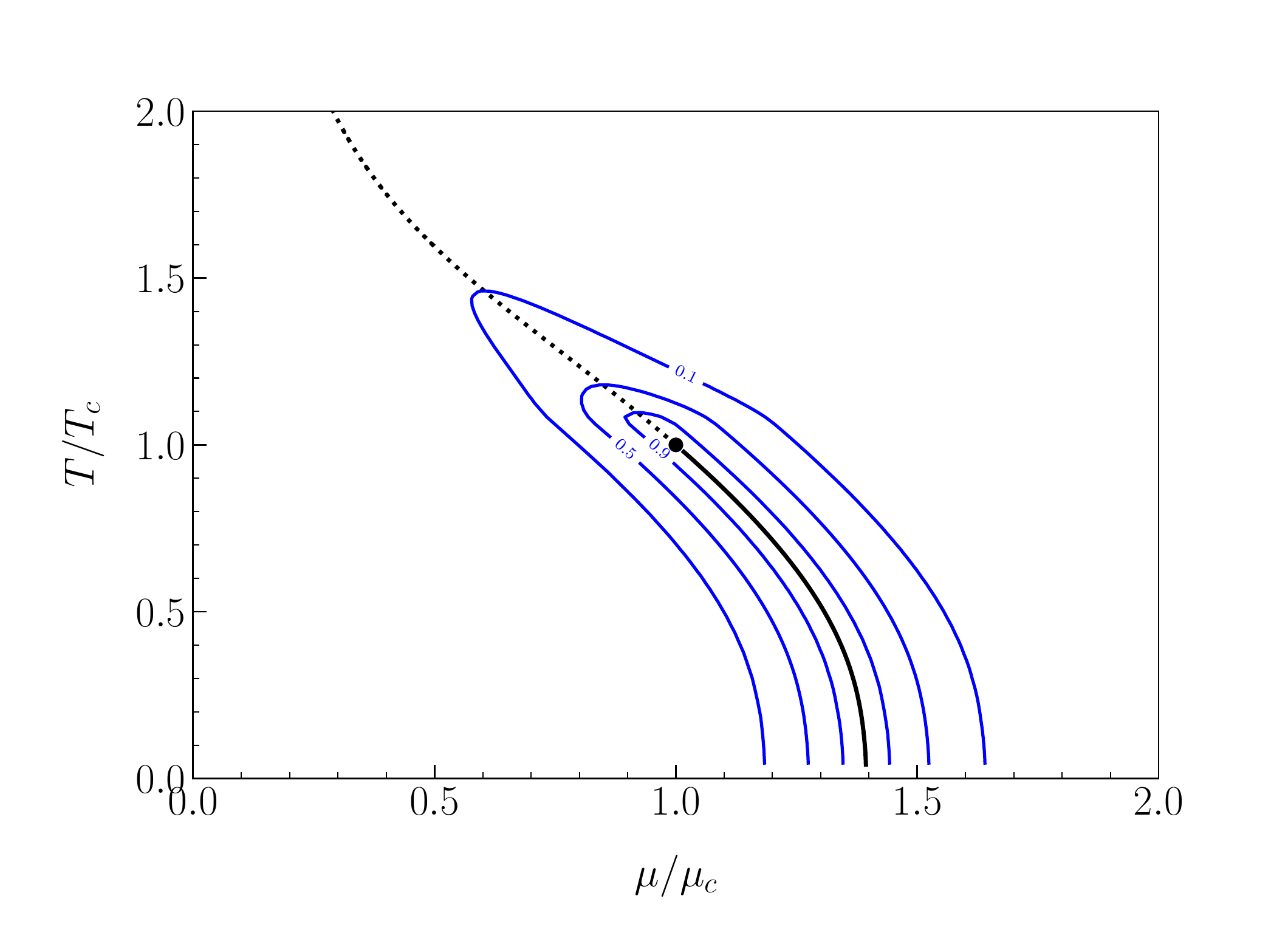}
\vspace{-10pt}
\caption{Window function contours with the factor of (\ref{fac}).}
\label{W2}
\end{figure}

It is interesting and worthwhile to plot contours of the window function in the $T$ versus $\mu$ plane.  Figure \ref{W1} shows contours of 0.9, 0.5, and 0.1.  
These extend to arbitrarily high temperature because Eq. \ref{Worig} depends on distance from the curve of $\mu_x(T)$.  Figure \ref{W2} shows the same contours but with the inclusion of the factor (\ref{fac}) which decreases the window function and monotonically reduces the contribution from the critical part of the equation of state above $T_c$. 

\section{Conclusions}
\label{conclude}

In this paper we proposed a way to merge the Schofield critical equation of state with a smooth background equations of state which has a smooth crossover from hadrons to quarks and gluons.  The method is generic and can be done with any physically reasonable background equation of state.  If there is a critical point, one can go around it without crossing the line of first order phase transition.  This means that there is a residue of the hadronic equation of state on the high density side and a residue of the quark-gluon equation of state on the low density side of that line.  Like the approach taken in Ref. \cite{Taylor1} our results reported here match smoothly onto lattice QCD simulations at $\mu = 0$.  The advantage of the approach taken here is that the equation of state is not limited to small values of the chemical potential because it is not based on an expansion in terms of $\mu/T$ for the background.  It also has the advantage of having a coexistence curve which has a symmetric inverted U shape in the $T$ versus $n$ plane.  The appendix points out that the approach taken in Refs. \cite{attract,Taylor1}, which relates Ising model and liquid-gas phase transition variables in a way different than ours, yields an order parameter with critical exponent $\beta$ which is not the baryon density but a linear combination of baryon and entropy densities.  The ``Schofield approach" and the novel approach taken in Ref. \cite{ourletter} are alternative ways of embedding critical behavior in a smooth background equation of state.  The goal, of course, is to use such equations of state in hydrodynamic simulations of heavy ion collisions in order to infer whether there is critical behavior.  They can also be used in numerical simulations of neutron star mergers where significantly high energy densities are expected to be achieved.

Further challenges, such as including not just the chemical potential for baryon number but also for electric charge and strangeness, and using a background equation of state which includes more realistic attractive and repulsive nuclear interactions at low temperature, will be explored elsewhere.

\section*{Acknowledgments}
This work was supported by the U.S. DOE Grant No. DE-FG02-87ER40328.  We thank the following for constructive comments on the manuscipt: Marlene Nahrgang, Jacquelyn Noronha-Hostler, Paolo Parotto, Christopher Plumberg, Krishna Rajagopal, Claudia Ratti, Thomas Sch\"afer, and Misha Stephanov.

\section*{Appendix}

To translate Ising model variables into QCD variables Refs. \cite{attract,Taylor1} perform the rotation
\ba
\frac{T - T_c}{T_c} &=& w \left[ \rho \sin\alpha_1 R (1-\theta^2) + h_0 \sin\alpha_2 R^{\beta \delta} h(\theta) \right] \nonumber \\
\frac{\mu - \mu_c}{\mu_c} &=& -w \left[ \rho \cos\alpha_1 R (1-\theta^2) + h_0 \cos\alpha_2 R^{\beta \delta} h(\theta) \right]
\label{Chiho}
\ea
where $w$, $\rho$, $\alpha_1$, and $\alpha_2$ are constants.  The standard parametric scaling equation of state summarized in Sec. II has 
$w \rho \sin\alpha_1 = 1$, $\sin\alpha_2 = 0$, $\cos\alpha_1 = 0$, and $w \cos\alpha_2 = -1$.  The motivation for this rotation is that otherwise the coexistence curve would be $\mu_x(T) = \mu_c$ for all $T \le T_c$.  This is not what is observed in real atomic systems nor what is expected in QCD.  The reasonable assumption in those papers was that $\alpha_2 - \alpha_1 = 90^{\circ}$ so that the axes are perpendicular.  The specific choice made in Ref. \cite{Taylor1} was $\alpha_1 = 3.85^{\circ}$.

One consequence of this rotation is that the order parameter is not the density, but a specific linear combination of the density and entropy density \cite{Mermin}.  In the notation of that paper the order parameter for $T \le T_c$ is
\be
\Psi = (n - n_c) \left( \frac{\partial \mu}{\partial \zeta} \right)_{\tau} + (s - s_c) \left( \frac{\partial T}{\partial \zeta} \right)_{\tau}
\ee
where
\ba
\tau &=& R (1-\theta^2) \nonumber \\
\zeta &=& h_0 R^{\beta \delta} h(\theta) 
\ea
Then
\be
\Psi = w \left[ -\mu_c \cos\alpha_2 (n - n_c)  + T_c \sin\alpha_2(s - s_c) \right]
\ee
and
\be
|\Psi| \sim (-t)^{\beta}
\ee
along the coexistence curve.  This is a natural consequence of the rotation because $n$ and $\mu$ are conjugate variables, as are $s$ and $T$.  

The approach we follow is represented by Eqs. (\ref{TmuN}).  Near the critical point one can approximate $\mu_x(T)$ by $\mu_c + \mu_x'(T_c) (T - T_c)$ where
 $\mu_x'(T_c)$ is finite and negative.  Then
\ba
\frac{T - T_c}{T_c} &=& \tau \nonumber \\
\frac{\mu - \mu_c}{\mu_c} &\approx& \zeta + \frac{T_c \mu_x'(T_c) }{\mu_c} \tau
\label{ourapprox}
\ea
and so $\Psi = \mu_c (n - n_c)$.  Comparing Eqs. (\ref{Chiho}) and (\ref{ourapprox}) results in either $\alpha_2 = 0$ or $\alpha_2 = \pi$ near the critical point.  In either case
\be
\tan\alpha_1 = - \frac{\mu_c}{T_c} \frac{1}{\mu_x'(T_c)} > 0
\ee
with $\rho$ arbitrary.

\end{document}